\documentstyle[epsfig]{aipproc}

\newcommand{\invpb  }{\mbox{\rm pb$^{-1}$}}
\newcommand{\qsq    }{\mbox{$Q^{2}$}}
\newcommand{\psq    }{\mbox{$P^{2}$}}
\newcommand{\ft     }{\mbox{$F_{2}^{\gamma}$}}
\newcommand{\ftxq   }{\mbox{$\ft(x,\qsq)$}}
\newcommand{\avqsq  }{\mbox{$\langle\qsq\rangle$}}
\newcommand{\gev    }{\mbox{$\rm GeV$}}
\newcommand{\gevsq  }{\mbox{$\rm GeV^2$}}
\newcommand{\epem   }{\mbox{$\rm e^+e^-$}}
\newcommand{\gsg    }{\mbox{$\gamma^{\star}\gamma$}}
\newcommand{\eb     }{\mbox{$E_{\rm b}$}}
\newcommand{\etag   }{\mbox{$E_{\rm tag}$}}
\newcommand{\ttag   }{\mbox{$\theta_{\rm tag}$}}
\newcommand{\wvis   }{\mbox{$W_{\rm vis}$}}
\newcommand{\zn     }{\mbox{$\rm Z^0$}}
\newcommand{\kt     }{\mbox{$k_{\mathrm{t}}$}}
\newcommand{\znhad  }{\mbox{$\zn/\gamma^{\star}\rightarrow$ hadrons}}
\newcommand{\qq     }{\mbox{$\rm q\overline{q}$}}
\newcommand{\ggtau  }{\mbox{$\gamma^{\star}\gamma\rightarrow\tau^+\tau^-$}}
\newcommand{\nch    }{\mbox{$N_{\rm ch}$}}

\def\etal{{\it et al.}}

\begin{document}
\begin{flushright}
UCL/HEP 2000-08 \\ OPAL CR444 \\
\end{flushright}
\vspace{-7mm}
\title{A high-{\boldmath$Q^2$} measurement of the photon structure function 
{\boldmath\ft} at LEP2}
\author{Russell J. Taylor}
\address{Department of Physics and Astronomy, UCL,\\ 
Gower Street, London WC1E 6BT, United Kingdom}
\maketitle
\vspace{-5mm}
\begin{abstract}
  The photon structure function \ftxq\ has been measured at \avqsq\
  of 706~\gevsq\, using a sample of two-photon events with a scattered
  electron observed in the OPAL electromagnetic
  endcap calorimeter. The data were taken during the years 1997-1999,
  when LEP operated at \epem\ centre-of-mass energies ranging from 183 to 
  202 \gev, and correspond to an integrated luminosity of 424 \invpb.
  This analysis represents the highest \avqsq\ measurement of \ft\ made
  to date.
\end{abstract}
\section*{Introduction}
We present a measurement of the hadronic photon structure function \ftxq\
at a higher value of the average momentum transfer squared, \avqsq, than has 
ever previously been reported. The measurement of \ft\ is interesting 
because of its potential to test perturbative QCD \cite{ref:F2QCD,ref:Nisius}.
In the high-\qsq\ domain the perturbatively calculable point-like contribution
to \ft, which rises logarithmically with \qsq, dominates over the 
non-perturbative hadron-like part.\par
The structure function \ft\ has been measured at \avqsq\ of 706 \gevsq\, 
using a sample of single-tagged two-photon events recorded by the OPAL 
detector between 1997 and 1999. These events (also referred to as \gsg\ 
events) can be regarded as deep inelastic scattering of an electron on a 
quasi-real photon, and the flux of quasi-real photons can be calculated 
using the equivalent photon approximation \cite{ref:Weiz}.\par
To study \ftxq\ the distribution of events in $x$ and \qsq\ is needed. These 
variables are related to experimentally measurable quantities by
\begin{equation}
  \qsq = 2\,\eb\,\etag\,(1- \cos\ttag)
\end{equation}
and
\begin{equation}
  x = \frac{\qsq}{\qsq+W^2+\psq}\,,
  \label{eqn:Xcalc}
\end{equation}
where \eb\ is the energy of the beam electron, \etag\ and \ttag\ are the 
energy and polar angle of the deeply inelastically scattered electron, 
$W^2$ is the invariant mass squared of the hadronic final state and 
$\psq=-p^2$, where $p$ is the four-momentum of the quasi real target photon.
The requirement that the associated electron is not visible in the detector 
ensures that $\psq \ll \qsq$, so \psq\ can be neglected when calculating 
$x$ from Equation~\ref{eqn:Xcalc}.
\section*{Data selection}
 This analysis uses data from the 1997 to 1999 LEP runs, with \epem\
 centre-of-mass energies ranging from 183 to 202 \gev. 
 The total integrated \epem\ luminosity is $424~\invpb$. 
 Candidate $\gsg \rightarrow \mbox{hadrons}$ events are required to satisfy
 the following selection criteria, in addition to several technical cuts
 to ensure good detector status and track quality.
\begin{enumerate}
\item A tagged electron is required; that is, a cluster in the OPAL 
  electromagnetic endcap calorimeter with an energy of at least $0.6\eb$ 
  and a polar angle $\theta$ in the range 230--500~mrad with respect to 
  either beam direction.
\item The energy of the most energetic electromagnetic cluster in the 
  hemisphere opposite to that which contains the tagged electron must 
  be less than $0.25\eb$.
\item The number of tracks originating from the hadronic final state
  must be at least 3.
\item The visible invariant mass \wvis\ of the hadronic system is 
  required to be in the range
  $2.5~\gev \le \wvis \le 50~\gev$.
\item The number of objects (tracks plus unassociated clusters), 
  belonging to the hadronic final state must be at least 9.
\item The energy deposited in a cone of 200~mrad half-angle about the 
  direction of the tag, excluding the tag itself, must not be more than
  2 \gev.
\end{enumerate}
\par
 Cuts 1--4 select a sample of candidate single-tag hadronic two-photon events,
 with double-tag events excluded by cut 2.
 Events with leptonic final states are rejected by cuts 3 and 5. The invariant
 mass cuts have two functions. The lower limit removes the low-$W$ region
 which is dominated by resonance production and is very difficult to model 
 accurately. The upper limit rejects background events from hadronic decays
 of \zn\ bosons, as does cut 6.\par
 A total of 348 events pass these cuts, with the data covering the range
 $270~\gevsq < \qsq < 2200~\gevsq$. There is a two sigma difference
 between the number of events selected in the 1997/8 data and that recorded 
 in 1999, with the 1999 data lying below the Monte Carlo expectation, 
 particularly at low \wvis. No explanation has been found for this. The 
 larger samples of \gsg\ events with the electron tagged in subdetectors 
 at lower polar angles are consistent between the two periods, suggesting 
 that the observed difference could well be purely statistical.
 However, as a precaution, the difference is included in the 
 systematic error in this preliminary analysis.\par
 The OPAL LEP1 analysis of \ft\ using tags in the same subdetector 
 \cite{ref:PR185} found the trigger efficiency to be 100\%. The present 
 analysis uses a tighter set of cuts, thus no inefficiency is to be expected,
 and a trigger efficiency of 100\% is assumed.
\section*{Monte Carlo modelling and background}
Monte Carlo programs are used to simulate \gsg\ events and to provide 
background estimates. 
The Monte Carlo generator used to simulate signal \gsg\ multiperipheral events
is HERWIG 5.9+\kt(dyn) \cite{ref:Herwig}. The GRV LO~\cite{ref:GRV} 
parameterisation of \ft\ was used as the input structure function.\par
The dominant background comes from the reaction $\znhad$. Also significant 
are non-multiperipheral four-fermion processes with \epem\qq\ final states
and the QED process \ggtau.
Less severe sources of background are estimated to account for
around 1\% of the data sample.\par
\begin{figure}[!tb]
\begin{center}
\epsfig{file=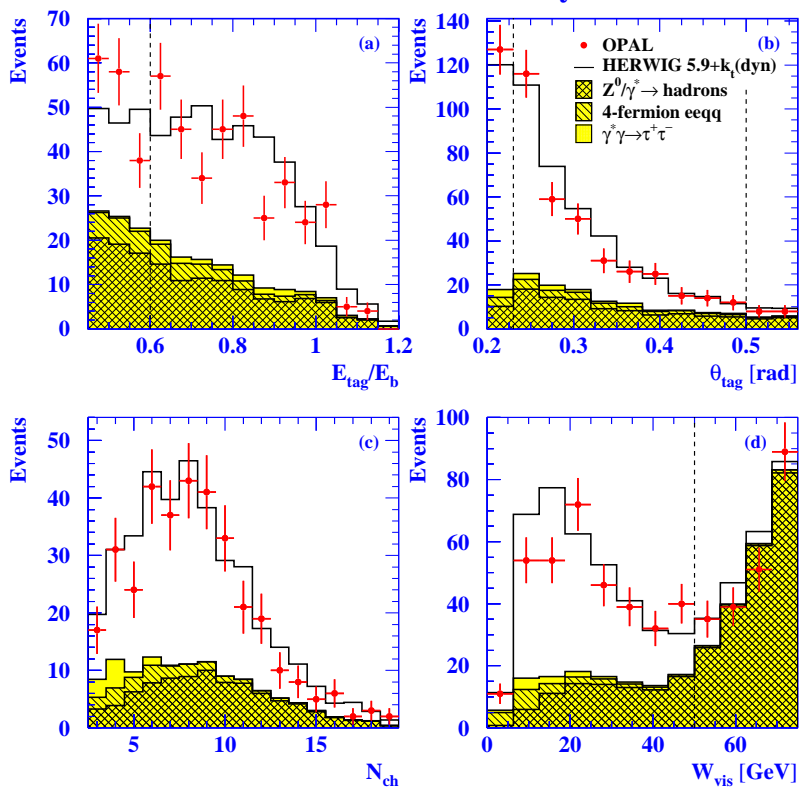,height=11.8cm}
\caption{Comparison of data distributions with the Monte Carlo prediction. The 
  open histogram is the sum of the HERWIG 5.9+\kt(dyn) prediction and the 
  contributions of the major background sources (shown
  as shaded histograms). 
         All selection cuts have been applied, except
         for any cut on the variable in the plot
         (indicated as dashed lines).
         The distributions shown are:
         (a) \etag/\eb, the energy of the tagged electron as a fraction
            of the beam energy, 
         (b) \ttag, the polar angle of the tagged electron,
         (c) \nch, the number of tracks originating from the hadronic
             final state, and
         (d) \wvis, the measured invariant mass of the hadronic final state.
         }
\label{datadists}
\end{center}
\end{figure}

Figure \ref{datadists} shows comparisons between data and
Monte Carlo distributions.
Figure~\ref{datadists}(b) shows the polar angle of the tagged electron. It 
can be seen that the Monte Carlo is somewhat higher than the data in the
polar range 260--350~mrad. Turning to variables describing the hadronic
final state, it can be seen that the number of charged tracks is reasonably
well described, Figure~\ref{datadists}(c), but that the Monte Carlo lies
above the data at low \wvis\ in Figure~\ref{datadists}(d) - which
correlates with high $x$.\par
\section*{Determination of \boldmath \ft}
The perennial problem in measurements of \ft\ is that, because the \gsg\
centre-of-mass system does not coincide with the laboratory system, 
the hadronic
final state, which must be measured to determine $W$, is only partially
observed in the detector. This leads to a dependence of the \ft\ measurement
on the Monte Carlo modelling, which is needed for the unfolding process
used to relate the visible distributions to the underlying $x$ distribution.
\par
\begin{figure}[tb]
\begin{center}
\epsfig{file=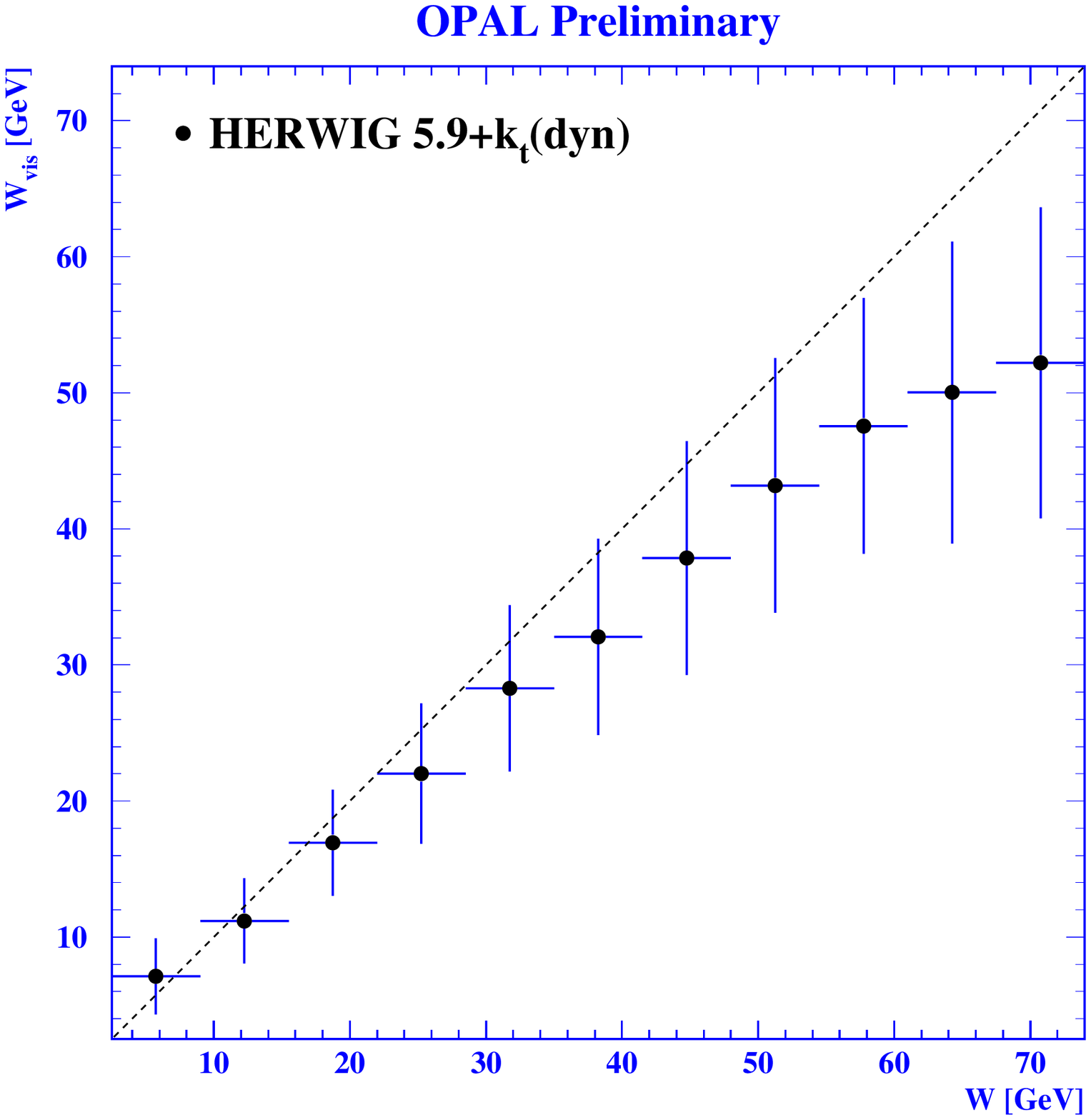,height=9.5cm}
\caption{The correlation between the generated hadronic invariant mass and
the measured value, as given by HERWIG 5.9+\kt(dyn). The vertical error 
bars represent the spread within each bin. The dashed line corresponds to
perfect correlation.
         }
\label{wrec}
\end{center}
\end{figure}

In the high-\qsq\ measurement presented here, however, the situation is not 
as serious as at lower 
values of \qsq. Because of the larger tagging angle, the hadronic final state
has much more transverse
momentum and as a consequence is better contained in the detector. 
Figure \ref{wrec} shows the correlation between the measured invariant
mass \wvis\ and the generated $W$ as given by HERWIG 5.9+\kt(dyn). 
It can be seen that the correlation is maintained throughout. This is in 
contrast with the situation observed in the lower \qsq\ analysis 
\cite{ref:PR314} where the correlation deteriorates at high W. 
As a consequence of this the result can be expected to have
a smaller dependence on the Monte Carlo modelling of the hadronic final
state.\par
After subtraction of background, the data are unfolded on a linear scale in
$x$ in the range $0.1 \le x \le 0.98$ using the GURU program~\cite{ref:GURU}.
\begin{figure}[tb]
\begin{center}
\epsfig{file=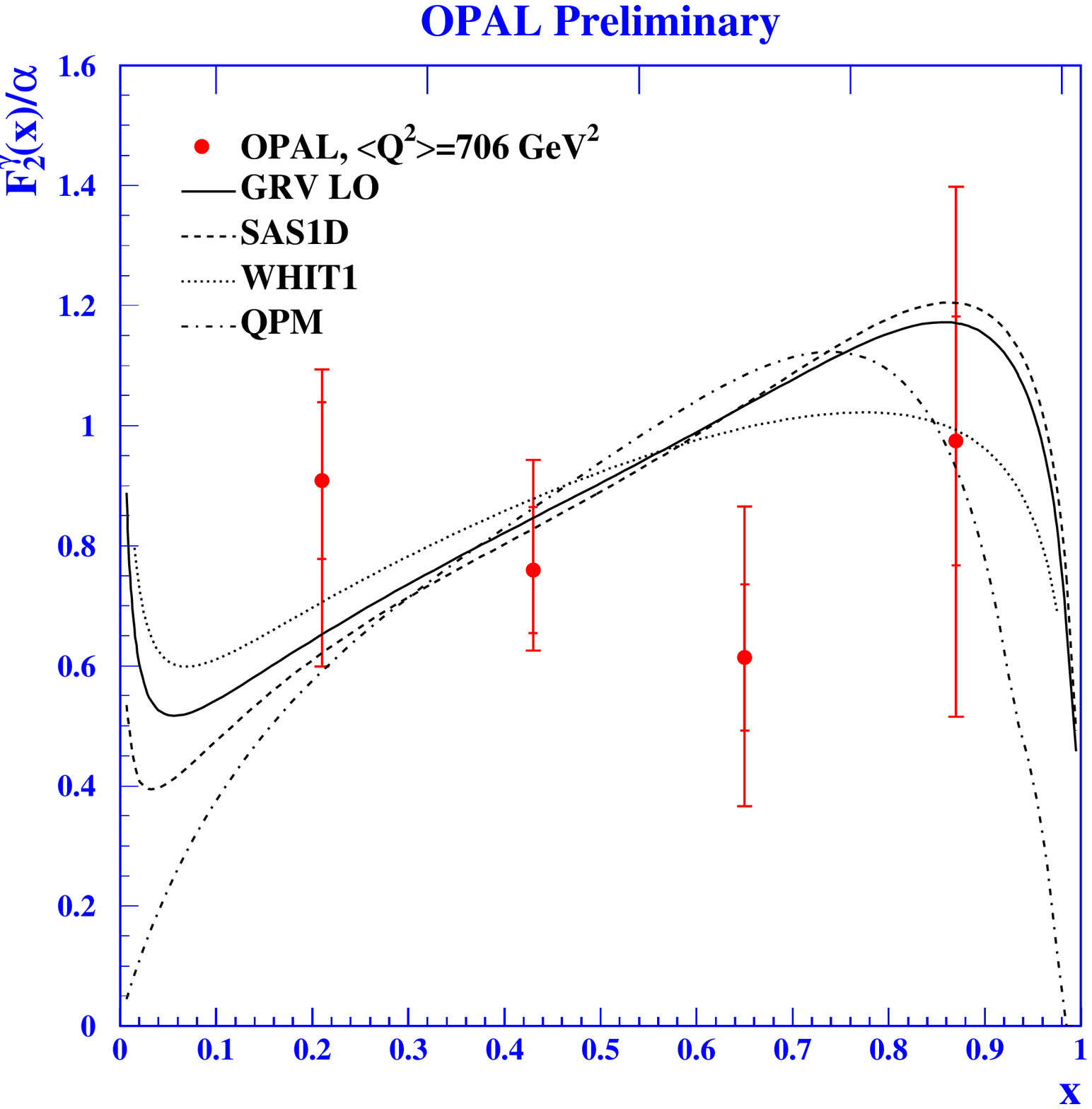,height=11.5cm}
\caption{The measured \ft\ at \avqsq\ = 706 \gevsq. The inner bars indicate 
the statistical error and the full bars the total error. 
The bin boundaries are indicated by the vertical lines at the top of the 
figure. 
The curves show the predictions of the GRV LO, SaS1d, WHIT1 and QPM structure 
functions, all for the \avqsq\ value of the sample.}
\label{result}
\end{center}
\end{figure}
Each data point is corrected for radiative effects and bin-centre corrections
are applied.
In Figure \ref{result} the data are compared to several theoretical 
calculations. The leading order parameterisations of \ft\ from GRV, 
SaS1d \cite{ref:sas1d} and WHIT1 \cite{ref:WHIT}, which all include a 
contribution from massive charm quarks, are described in detail in 
reference \cite{ref:Nisius}. The naive quark-parton model (QPM) simulates 
only the point-like component of \ft, and is calculated for four active 
flavours with masses of 0.325~\gev\ for light quarks and 1.5~\gev\ for charm 
quarks. It can be seen that in this high-\qsq\ regime the differences between
the models are relatively small, particularly in the central $x$-region. 
The differences between the QPM and the other models are much smaller than 
at lower \qsq, where the photon has been shown to have a significant
hadron-like component \cite{ref:PR314}. All the predictions are compatible 
with the data in three of the $x$ bins, but overshoot the data in one bin.
\section*{Conclusions}
The photon structure function \ft\ has been measured using deep inelastic
electron-photon scattering events recorded by the OPAL detector during the
years 1997--1999. The \avqsq\ value of 706 \gevsq\ represents the highest
measured thus far. \ft\ has now been measured by OPAL at \avqsq\ values 
ranging from 1.9--706 \gevsq.\par

\end{document}